\documentclass[conference]{IEEEtran}

\usepackage{algorithm}
\usepackage{algpseudocode}
\usepackage{algpseudocode,algorithm,algorithmicx}
\usepackage{textcomp}
\usepackage[english]{babel}
\usepackage[utf8x]{inputenc}
\usepackage{amsmath}
\usepackage{amssymb}
\usepackage{graphicx}
\usepackage{subcaption}
\usepackage[export]{adjustbox}
\usepackage{array,multirow}
\usepackage[table]{xcolor}
\usepackage[colorinlistoftodos]{todonotes}
\usepackage[normalem]{ulem} 
\usepackage{xcolor}
\usepackage{fancyvrb}
\usepackage{wrapfig}
\usepackage{tikz}
\usepackage{enumitem}

\usepackage{listings}

\newcommand{\code}[1]{{\lstinline!#1!}}

\author
{
  	\IEEEauthorblockN{Rokas Volkovas}
  	\IEEEauthorblockA{QMUL\\
  	London, UK\\
  	r.volkovas@qmul.ac.uk}
    \and
  	\IEEEauthorblockN{Michael Fairbank}
  	\IEEEauthorblockA{University of Essex\\
  	Colchester, UK\\
  	m.fairbank@essex.ac.uk
  	}
    \and
  	\IEEEauthorblockN{John Woodward}
  	\IEEEauthorblockA{QMUL\\
  	London, UK\\
  	j.woodward@qmul.ac.uk}
    \and
  	\IEEEauthorblockN{Simon Lucas}
  	\IEEEauthorblockA{QMUL\\
  	London, UK\\
  	simon.lucas@qmul.ac.uk}
}

\begin{document}
\title{Mek: Mechanics Prototyping Tool for 2D Tile-Based Turn-Based Deterministic Games}

\maketitle
\thispagestyle{plain}
\pagestyle{plain}

\begin{abstract}
There are few digital tools to help designers create game mechanics. A general language to express game mechanics is necessary for rapid game design iteration. The first iteration of a mechanics-focused language, together with its interfacing tool, are introduced in this paper. The language is restricted to two-dimensional, turn-based, tile-based, deterministic, complete-information games. The tool is compared to the existing alternatives for game mechanics prototyping and shown to be capable of succinctly implementing a range of well-known game mechanics.
\end{abstract}

\IEEEpeerreviewmaketitle

\section{Introduction}
Existing research in tool-assisted-game-design is primarily dedicated to the development of game generation algorithms. Frameworks such as GVGAI~\cite{perez2016general} are determined to provide a groundwork for the development of general AI or complete game generation automation. As such, the languages used for the game specification are flexible enough to describe games with significantly different player goals, but too rigid to be used for describing new interactions. 

The most flexible tool for a digital game designer remains a general purpose programming language, which requires a good knowledge of programming and is a significant barrier of entry. This is not the case with other disciplines in game development, such as visual art or sound design. To begin improving the situation towards one where game designers have an environment -- equivalent to the environment of what \texttt{Photoshop} is to visual artists -- a language, abstracting the common tedious details of digital games is introduced here. 

\texttt{MekLang}, the introduced language, sacrifices some description generality for its uniformity through enforcing the mechanics to fall into the two-dimensional (2D), turn-based, tile-based, complete-information categories. In contrast to other existing attempts, the language emphasizes development of mechanics, rather than games.

\subsection{Games vs Mechanics}
For a more precise discussion of the details and the importance of mechanics, it is necessary to draw a clear line between what is meant by \texttt{games} and \texttt{mechanics} within this document. A mechanic is a state transition definition. In other words, a mechanic describes how the game is allowed to change over time or with every move available. A game, on the other hand, is a set of mechanics with an end state -- a state where no moves can be made.

Consider an infinite \texttt{Chess} game board. The way the bishop moves is a mechanic. Specifically, the ability to move the piece anywhere on the board, as long as it is within a straight diagonal line from the current position, is the mechanic. Similarly, adding a knight on the same infinite board, increases the number of mechanics to three: movement of the bishop, movement of the knight and their interaction with each other. That is, a combination of mechanics is also a mechanic. The limitation of the board to an \texttt{8x8} grid introduces the mechanic of pieces not being able to move off the board. Mechanics by themselves are not games. Adding an end state -- which is also a mechanic -- is what makes the complete set of mechanics a game.

\begin{table*}[t]
\centering
\begin{tabular}{ | l | c | c | c | c | c | }
  \hline
    Tool        & Language  & Players   &  Game Creation    & Goal                          & Restrictions\\ \hline
    Mek         & Visual    & N/A      &  No               & Game Mechanics Prototyping   & DD\\ \hline
    Machinations& Visual    & N/A       &  No               & Game Systems Design           & Abstract\\ \hline
    Ludi        & Text      & 2         &  Automated        & Automate Board Game Design    & 2P, DD\\ \hline
    ANGELINA    & Text      & 1         &  Automated        & Automate Game Creation        & Depends on version\\ \hline
    GVGAI       & Text      & 1-2       &  Manual \& Auto   & Testbed for General AI        & 2D Sprite-based\\ \hline
\end{tabular}
\caption{Game Design Tool Comparison}
\label{tab:comparison}
\end{table*}

\subsection{Analysis Implications}
The explicit separation of the game and its mechanics is important for both design and analysis purposes. In Chess, a player wins the game by check-mating the opponent's King (making it unable to escape a check). Now consider a Chess variant where the player wins by getting her own King check-mated. These game versions can be argued to be significantly different in terms of how the players will approach playing them. Hypothetically, assuming there is a perfect artificial measure of fun, one game might be found to be fun and the other as far away from fun as possible. The only mechanic that was changed was the end condition. In this situation, no practical information is gained about the inner workings of the game, other than its overall quality. In other words, the game designer learns nothing about the mechanics, the building blocks of the game. 

Here, we argue, that mechanics can and should be separated from games and analyzed as distinct units of game design. Much like in art, a teacher can tell whether or where the basic shapes a student is drawing are well presented, one should be able to tell whether a mechanic, in isolation, is fit for the purpose it is being implemented for. Mechanics build up games like stepping stones, and they alone can give solid indication on how the game will function. Structured mechanic analysis will provide a groundwork for experimentation through rapid prototyping and is most sensible when the mechanics are defined in a mechanics-focused language. Section~\ref{sec:lang} describes the first incarnation of such a language, created for the purpose of prototyping mechanics.

\section{Existing Work}
\label{sec:lit}
This section explores the inner-workings of the alternative tools, identifying their strengths and weaknesses for mechanics prototyping. Table~\ref{tab:comparison} compares the features of \texttt{Mek} with those the most prominent existing game design tools. Notes:
\begin{itemize}
  \setlength\itemsep{0.05em}
    \item DD = Discrete, Deterministic
    \item Discrete = turn-based, tile-based
    \item All languages are custom (not build directly on another)
    \item \emph{Players} is the number of players supported in games
    \item \emph{ANGELINA} creates different genre games per version
    \item \emph{Machinations} helps designing conceptual systems
\end{itemize} 
MekLang is conceptually most similar to the \emph{Machinations} tool in terms of its purpose. The other tools listed use game design as a fertile ground for AI experimentation (\emph{Ludi}, \emph{ANGELINA} and \emph{GVGAI}). There exist other tools that can be use for game design by proxy, such as \emph{PuzzleScript}, \emph{RPGMaker} or any other engine geared towards creating games. However, these tools are ignored as not being focused on advancing game design explicitly. 

\subsection{Machinations}
Joris Dormans' \emph{Machinations}~\cite{dormans2009machinations}\cite{machisutra} tool can easily be considered the most formalized design tool implemented to date. It has the look of UML diagrams. The tool focuses on the ability to express what is called game \emph{e conomies}, which are the transfer of game numerical values (resources) from one container to another over the course of the game. This focus bias allows the designer to quickly express high-level resource interactions common in most traditional board games with the ability to define value transitions using common mathematical operations. 

The language expressiveness power, however, comes at the cost of abstracting game details to sometimes unrecognizable game implementations -- not only does it separate the level design from the mechanics, but it removes level design entirely. The removal of level design allows game designers familiar with the system to discuss concepts quickly, however the design of even a relatively basic resource-focused game like \emph{Monopoly} depends heavily on the ``level" (board) representation. The lack of level design expressiveness also limits the ability of automation, both creating new mechanics and analyzing the existing ones, not knowing how they would interact in the full game. On the other hand, adding such capabilities seems to be within reasonable reach of the language, but without the source code availability or author's support this is uncertain. 

\subsection{Ludi}
\emph{Ludi}~\cite{browne2010evolutionary} is a framework devised specifically for the game mechanics delivery automation (the polar opposite of manual design). It focuses on finding \emph{fun} game rule-sets using an evolutionary-algorithm based system. \emph{Ludi} is a good example of the existing interest of analyzing mechanics. However, while it uses a language that could potentially be used by game designers to explore their designs, its definition is geared towards the automated design iteration abilities. The system was shown to even be capable of producing a game that was released commercially~\cite{browne2011yavalath}. 

The restrictions it imposes onto the created games are that they have to be turn-based, grid-based, deterministic, two-player and perfect information. The most significant of these is the two-player requirement. It restricts designers when considered in comparison to other mechanics or general languages. The rules are expressed using a text-based scripting language, which features the ability to define a set of distinct board types: traditional 8-neighbour square grid, hexagons, triangles and some less familiar ones. The focus of the language towards design automation is apparent when reading the described game definitions. The definitions are very flexible in how they can be written out, but may be hard to parse for someone not familiar with the language features in-depth. Nonetheless, the system shows what features lend themselves more easily to AI analysis and design metrics.

\subsection{ANGELINA}
\emph{ANGELINA} is an example of game design being approached with another distinct set of motivations. It is a system created to produce complete games autonomously, using procedural generation creatively (e.g. looking up images based on text)  and has gone through a number of iterations, with the latest one~\cite{cook2014ludus} making 1990s dungeon crawling RPG-like 3D games. 

In stark contrast to the systems described above, it attempts to entirely remove the designer from game development. While at first glance, this would rule it out from being valid consideration or a comparison to tools aimed to aid game designers, it is important to recognize that as the field becomes more and more automated, the approaches made towards automation help gain insight into what features to include or avoid in order to make even designer-centric systems more expandable. From this point of view, it is useful to note how \emph{ANGELINA} handles game resource management, both using pre-designed pieces for levels and objects combined with evolution. The system does not directly expand on any of the specific algorithms but tackles the issue of gluing them all together in a cohesive manner.

\subsection{GVGAI}
The General Video Game Artificial Intelligence (\emph{GVGAI}) framework~\cite{perez2016general} is a system created with the goal of encouraging AI researchers to focus their attention on general problem solving methods in order to allow them to be more readily transferable to new domains. Through defining a common interface to a large number of distinct games, the framework facilitates exploration of a number of game development areas, including: general AI agents playing unseen games~\cite{guerrero2017beyond}, generating levels~\cite{neufeld2015procedural} and, most relevantly, generating game rules~\cite{khalifa2017general}. 

The rule generation in its case is treated as a problem to be solved, given a specific level. The generation is done on the game description language used in \emph{GVGAI}, which is evolved from~\cite{ebner2013towards}. The language is succinct in terms of what it aims to represent, which is relatively complex games, but this feature comes at the cost of flexibility, as the rules specified rely on engine-available features. This approach does not align well with manual design of new mechanics. The framework is also directed more towards generality, which allows for creative research, such as mechanics recommender systems~\cite{machado2016shopping}, but might deter in-depth focus for better or worse. 

\subsection{Design Considerations}
The tools outlined in the previous paragraphs were found to be the most prominent examples of the state of the art tools available to game designers comparable to Mek. This is not to indicate that no other work has been done towards the same goal, only to showcase the work that's practically usable. A number of primarily philosophical considerations have also been discussed, including~\cite{smith2009computational}, which outlines a method of potentially more useful prototype iteration in their developed engine~\cite{smith2009prototyping}, encouraging both human and machine play-testing. Or~\cite{treanor2012game}, which describes a system to express game ideas and concepts instead of concrete designs. 

Any type of game-specific design tools, such as~\cite{guzdial2016learning}~\cite{khalifa2016general}, or even use of machine learning to define and combine mechanics from existing games~\cite{barros2018data}~\cite{guzdial2018automated} is excluded from consideration as being either not directly usable or too specific to be useful for game designers. Also not covered was material such as the Stanford GDL~\cite{genesereth2005general}, which could be considered a precursor to the language used in GVGAI framework.

\section{Language Requirements and Restrictions}
\label{sec:req}
Prior to describing the details of the language, it is important to understand the requirements outlined for it and their significance. Intuitively, every digital game can be described as a Markov Process, with every distinct frame of a game being a different state. However, this naive implementation yields impractically vast numbers of states and their transitions.
\subsection{Scalability}
While Markov Process representation may be practically applied to anything with few states, making changes is a tedious process and the implementation does not scale to larger variations of the same game/mechanic. Consider implementing an adjacently moving piece in \texttt{2x2} sized grid-world. This results in \texttt{4} possible states (indicating unit position) with \texttt{8} connections between the states (\texttt{2} adjacent connections from each tile). The setup grants complete flexibility over how the connections between the states are connected. 

Now consider expanding the grid-world size to \texttt{3x3}. The movement mechanic is identical, but the expansion requires adding \texttt{5} extra states, with \texttt{20} connections. The representation complexity (number of needed states and connections) increases quadratically with the grid size. This is undesirable. Thus the language should be capable of scaling mechanics to any board size.

\subsection{Discrete 2D and Information Completeness}
With the conceptual goal in place, the domain of game mechanics needs to be defined. The language is chosen to specialize in two-dimensional, turn-based, tile-based, complete-information games. These categories were chosen due to the ease of information presentation, computational efficiency and large range of distinct existing games falling under these categories. That is, these restrictions strike a favourable balance between the implementation complexity and mechanic definition flexibility. 

The following games -- and more importantly, their mechanics -- fall under the defined restrictions: \emph{Chess}, \emph{Snake}, \emph{Othello}, \emph{Tetris}, match-3 games (\emph{Bejeweled}, \emph{Candy Crush} etc.), \emph{Sudoku}, \emph{Sokoban}  and many others. These games are all mechanically distinct -- they have mechanics which are naturally associated  with those specific games, yet they all share the common features of having a limited size two-dimensional grid if square tiles, are played in turns and present all gameplay information to the player at all times.

\subsection{Mechanics Focus}
The purpose of the developed language is to make the design of mechanics more accessible. This core principle establishes that the language is \emph{not} for making games, but creating the mechanics. These mechanics can then be taken outside of the language confines to make full-games. 

The tools to support game development are not the same ones that support mechanics development. This may be difficult to grasp, since no widespread tools support mechanics development exist. Games need: mechanics, art, sound, general programming etc. To help the understanding, mechanics are to be looked at as what 3D models are to games -- individual pieces of art which can be presented by themselves and need some small (relative to the model creation) amount of work to be utilized in a game. If a model looks good in modelling software, then it will likely look good in game. This is not guaranteed though, due to the possibility of game-sensitive issues, such as clashing art-styles with other models. The same holds for game mechanics. Yet, there are no tools anywhere near as expressive as the state of the art modelling software. The ultimate goal here is to build one. The tool is a small step towards this idealistic goal.

\newpage
\section{Mek}
\label{sec:lang}
\begin{figure}[h]
\centering
\includegraphics[width=\linewidth]{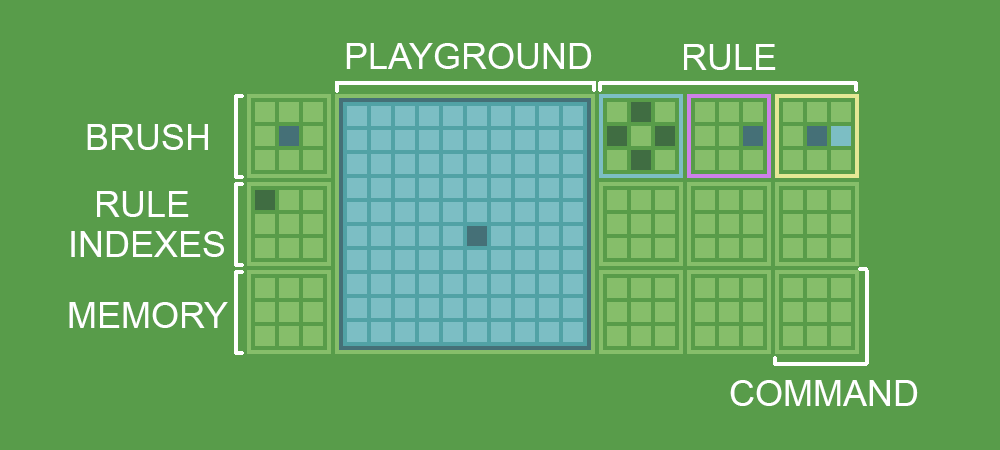}
\caption{Complete Mek interface}
\label{fig:mek}
\end{figure}

\emph{Mek} is the proposed visual mechanics prototyping tool. \emph{MekLang} is the language Mek uses to implement mechanics. Understanding MekLang is easier through first understanding Mek. Figure~\ref{fig:mek} shows how the complete interface of Mek looks like. The interface is a combination of a number of individual parts. The inner workings of all the interface parts and their interactions are described in this section.

\subsection{Playground}
\begin{figure}[h]
\centering
\includegraphics[width=\linewidth]{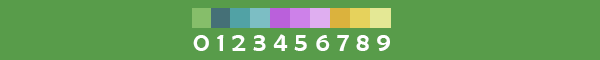}
\caption{Available playground tile colors with their indexes.}
\label{fig:colors}
\end{figure}

The playground is a \texttt{10x10} grid of tiles. In Figure~\ref{fig:mek}, it is seen as the large grid center-left of the image. This grid is used to represent the current board state, upon which the designer-implemented mechanics operate. A playground tile can be one of the \texttt{9} colors shown in Figure~\ref{fig:colors}. The colors are labeled with their respective color indexes.

\begin{figure}
\centering
\includegraphics[width=\linewidth]{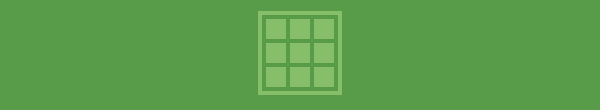}
\caption{Command: a \texttt{3x3} grid of tiles.}
\label{fig:command}
\end{figure}

\subsection{Command}
A command is a \texttt{3x3} grid of tiles. Figure~\ref{fig:command} shows how one appears in Mek. This grid is used to define an operation to be done on the playground. To change the command behaviour, each one of the tiles can be clicked using the mouse. When clicked, the command tile changes color to the next available one, cycling back when the last color is reached. The number of available colors differs depending on the type of a command, but by default command tiles all look like ones in Figure~\ref{fig:command}. The type of a command is indicated visually by the color of the outline of the grid. Table~\ref{tab:comTypes} shows all the existing primary command types and their associated colors. The behaviour associated with each command and their variations are specified in Section~\ref{sec:meklang}.
\begin{table}[b]
    \centering
    \begin{tabular}{|r|c|c|}
        \hline
        Command Type        & Outline Color  & Tile Color Set \\ \hline
        \texttt{WRITE}      & Yellow         & All, except dark green \\ \hline
        \texttt{CHECK}      & Light Purple   & All \\ \hline
        \texttt{SHIFT}      & Dark Green     & Dark green only \\ \hline
        \texttt{ROTATE}     & Light Blue     & All \\ \hline
        \texttt{CYCLE}      & Orange         & All, except dark green \\ \hline
        \texttt{CALL}       & Dark Blue      & Dark green only  \\ \hline
    \end{tabular}
    \caption{List of all the primary command types and their associated outline colors and available tile color sets.}
    \label{tab:comTypes}
\end{table}

\subsection{Command cycling}
The color cycling operation happening on command tile click is equivalent to the following modulo operation:
\[ ColorIndex_{n} = (ColorIndex+1) \% |AvailableColors| \]

\begin{figure}[h]
\centering
\includegraphics[width=\linewidth]{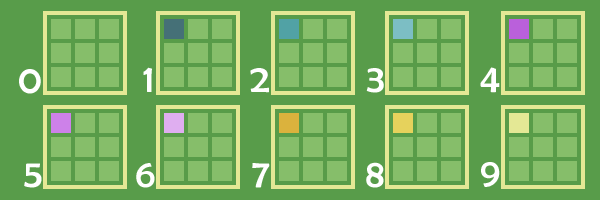}
\caption{Command color cycling. From \texttt{0} to \texttt{9}: command grid transformation after each top left tile click.}
\label{fig:cycling}
\end{figure}

Figure~\ref{fig:cycling} shows the sequence of enumerated single command states. Specifically, the command labeled \texttt{0} is the default state of the \texttt{WRITE}-type command (the type is indicated by the yellow border color). Once the top-left tile is left-clicked, the command transforms into the one labeled \texttt{1}. Left-clicking the top-left tile again transforms the command into the one labeled \texttt{2}. When the command state labeled \texttt{9} is reached, the next left-click on the same tile leads back to the grid labeled \texttt{1}. If the tile is right-clicked, the states in the image are traversed in the opposite direction (\texttt{2} to \texttt{1}, \texttt{1}, to \texttt{9}, \texttt{9} to \texttt{8} etc.). Middle-clicking the tile, resets it back to the color seen in state \texttt{0}. This describes the color cycling for the command of type \texttt{WRITE}. Other commands might have less colors available. Unavailable colors are skipped.

\subsection{Command Execution}
\begin{figure}[h]
\centering
\includegraphics[width=\linewidth]{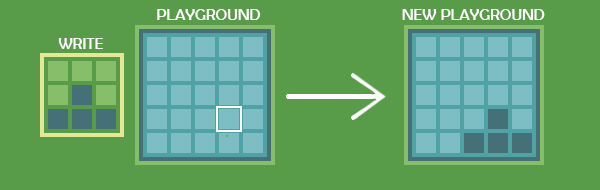}
\caption{\texttt{WRITE} command execution: when the tile highlighted in white is clicked, the left playground transforms into the one on the right.}
\label{fig:write}
\end{figure}

A command can be executed. Executing a command means performing an action on the board (playground + memory, introduced later) state. This action may or may not transform the board state. The nature of the action depends on the type of the command. For example, execution of a \texttt{WRITE} command, would cause certain tiles on the playground be overwritten with colors of the ones contained within the command, which is showcased in Figure~\ref{fig:write}. The \texttt{WRITE} command with (the \texttt{5x5} version of) the playground is shown the left. When the tile highlighted in white  is clicked, the playground becomes the one shown on the right. The execution of the command meant copying the non-light green contents of the command grid to the playground, relative to the clicked tile location.

\newpage

\subsection{Rule}
\begin{figure}[h]
\centering
\includegraphics[width=\linewidth]{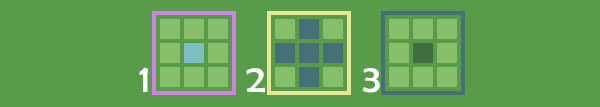}
\caption{Rule: a list of commands. The command outlines indicate their types: 1. \texttt{CHECK} 2. \texttt{WRITE} 3. \texttt{CALL}}
\label{fig:rule}
\end{figure}

A rule is a list of commands. The order in which the commands interact with the board state depends on its rule position. Figure~\ref{fig:rule} shows how a \texttt{3}-command rule would appear in Mek. In the figure, there are three commands each with a different type (indicated by the different outline colors). The rule can be executed. Executing a rule means executing its commands in order, one by one. Some commands may terminate the rule execution. In this case the remaining commands are not executed.

\begin{figure}
\centering
\includegraphics[width=\linewidth]{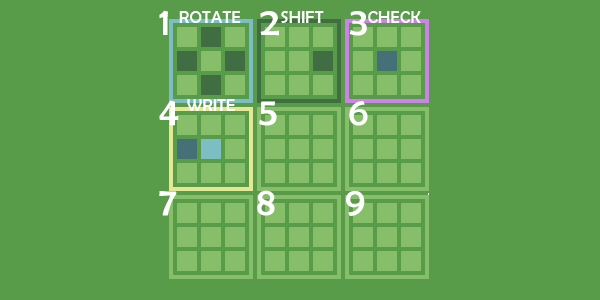}
\caption{A rule in Mek: a list of \texttt{9} commands.}
\label{fig:mekrule}
\end{figure}

In Mek, rules are \texttt{9} commands long. Figure~\ref{fig:mekrule} shows how a rule looks like in Mek. By default, all rule commands are empty (light green command outline). The number labels indicate the order in which they are executed. In the figure, the commands \texttt{5} to \texttt{9} are empty. Empty rules are skipped when a rule is executed. Listing~\ref{ls:rule} shows the pseudo-code equivalent of the rule logic shown in Figure~\ref{fig:mekrule}. 

\vspace{10pt}
\begin{lstlisting}[label={ls:rule}, caption={Example rule execution pseudo-code. Lines starting with \# indicate the beginning specific command type functionality.},captionpos=b]
func execute_rule(focus):
    #ROTATE
    for rotation in [0, 2, 4, 6]:
        focus.save()
        focus.rotate(rotation)
        #SHIFT
        focus.shift(RIGHT)
        #CHECK
        var tile = focus.tiles[CENTER]
        if tile.color = DARK_BLUE: 
            #WRITE
            focus.tiles[CENTER].color = L_BLUE
            focus.tiles[LEFT].color = D_BLUE
        focus.reset()
\end{lstlisting}
The pseudo-code makes use of the focus concept, explained in Section~\ref{sec:focus}. In the pseudo-code, the focus is an object, containing a reference to the tiles of the playground the commands are being executed upon.

\subsection{MekLang Process}
MekLang is the language Mek uses to implement the mechanics. The mechanics are built out of rules and commands. The mechanics operate on the playground. At its simplest form, MekLang in Mek works like this:
\begin{enumerate}
    \item A tile on the playground is clicked
    \item The available rules are executed
    \item The process repeats
\end{enumerate}




\subsection{Focus}
\label{sec:focus}
The \texttt{3x3} grid of playground tiles each command operates upon are said to have the \emph{Focus} on them. The focus has two parameters: position and rotation. By default, before command execution, the focus center location is set to match the location of the clicked tile. The focus position, as well as the rotation, can both be changed by specific commands, which are described later. Moving the focus affects which tiles are seen by the executed commands, whereas rotating the focus affects how they are seen. 

\subsection{Rule Index Grid}
\begin{figure}[h]
\centering
\includegraphics[width=\linewidth]{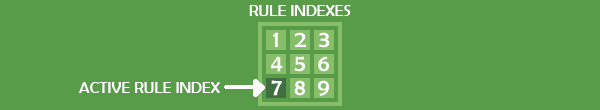}
\caption{Rule index grid: the dark green colored tile indicates the currently visible rule index.}
\label{fig:rule_indexes}
\end{figure}

One rule may not be enough for more complex mechanics. To combat this, Mek has \texttt{9} rules available. However, there is not enough screen space to show them all at the same time. To view and modify the different rules, the rule index grid is used. The rule index grid is shown in Figure~\ref{fig:rule_indexes}. The shown tiles are indexed by the rule numbers they lead to. Clicking the tile changes the visual rule commands to the ones of the selected rule. This is only a visual change for ease of interaction.

\newpage
\subsection{Tile Toggling Example}
\begin{figure}[h]
\centering
\includegraphics[width=\linewidth]{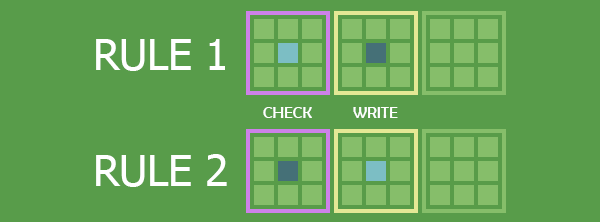}
\caption{Toggle mechanic rules: \texttt{CHECK} commands followed by \texttt{WRITE} commands of opposite color.}
\label{fig:togrule}
\end{figure}

A small step above in terms of complexity in comparison to the tile coloring example is the tile toggling mechanic: 
\begin{itemize}
    \item when a tile is clicked, toggle its color
    \begin{itemize}
        \item toggling a light blue tile turns it dark blue
        \item toggling a dark blue tile turns it light blue
    \end{itemize}
\end{itemize}
The simplest way to implement the mechanic is to use two rules, one for each toggle scenario. Figure~\ref{fig:togrule} shows how the first three commands of the two rules would look like. When a tile is clicked, both rules are executed in order.

The figure utilizes the already used but yet to be defined command, which is the \texttt{CHECK} command (purple command border). This command compares the values of the tiles focused on with the ones in the command. If the tiles mismatch, the rule being executed terminates. These rules implement the toggle mechanic.

\subsection{Implementation of Existing Games}
The examples presented above are a useful when describing MekLang, but are still not representative of the language's flexibility. The language was used to implement a number of well known games (or at least their core mechanics). Their representation complexity, in terms of the number of rules and commands used within the implementation, in MekLang is shown in Table~\ref{tab:games}. 

Note, that since MekLang is capable of implementing \emph{Game of Life}, it is Turing Complete, which implies that given endless memory it is capable of implementing all computer programs. However, this feature does not mean that doing so is simple enough in comparison to the alternatives. On the other hand, the number of commands and rules used to implement the different games arguably should give a solid indication of the language capabilities. 

\begin{table}[h]
\centering
\begin{tabular}{|r|c|c|}
\hline
Game                                & Rules & Commands  \\\hline
Sliding Puzzle                      &   1   &   3       \\\hline
Nim (Subtraction)                   &   2   &   8       \\\hline
Solitaire (Marbles)                 &   3   &   9       \\\hline
Sokoban (Pushing)                   &   3   &   9       \\\hline
Game of Life                        &   4   &   16       \\\hline
Noughts and Crosses                 &   5   &   19       \\\hline
Othello / Reversi                   &   5   &   21      \\\hline
Connect 4                           &   6   &   26      \\\hline
\end{tabular}
\caption{Known game implementation complexity}
\label{tab:games}
\end{table}

\section{MekLang}
\label{sec:meklang}
To make the implementation of all the games listed in Table~\ref{tab:games} possible a number of specific command types were used. All the different commands, along with the additional behaviours of MekLang are covered in this section.

\subsection{Memory}
Memory is an additional \texttt{3x3} tile grid that the commands can operate on. It has all the same features that the playground does, except its size. Memory and playground together form the entire board state. The memory grid is seen in the bottom-left corner of Figure~\ref{fig:mek}.

\subsection{Modes of Operation}
There are two modes of operation in Mek. The modes are:
\begin{enumerate}
    \item \texttt{BRUSH}
    \item \texttt{NORMAL}
\end{enumerate}
The modes affect what rules are executed when a tile is clicked. In \texttt{BRUSH} mode, only one rule, with a single command is executed. The command type is set to \texttt{WRITE} and cannot be changed. The purpose of this mode is to be able to quickly change the board state without executing the implemented mechanics. The implemented mechanics are executed in the \texttt{NORMAL} mode. The modes can be switched between at any point a tile click is expected.

\subsection{Execution Scheduling}
When executed, some commands may complete their action immediately, or schedule it to be done after all the rules are executed. Immediately performed actions affect what the subsequent commands interact with; delayed actions do not. If multiple commands schedule overlapping actions, they are applied in the scheduling order.

\subsection{\texttt{WRITE}}
The simplest command is of the \texttt{WRITE} type. When a \texttt{WRITE} command is executed, the tiles at the focus location are overwritten with the contents of the command grid. Light green command grid tiles are ignored, thus keeping the original value of the target grid. It has a number of variations\footnote{Command variations have distinct shades of the base command color. They can be thought of as functions with different parameters.}. Including the default, the variations are:
\begin{itemize}
    \item \texttt{WRITE}
    \item \texttt{WRITE TO MEMORY}
    \item \texttt{WRITE FROM MEMORY}
    \item \texttt{WRITE FROM PLAYGROUND}
    \item \texttt{WRITE INSTANT}
    \item \texttt{WRITE TO MEMORY INSTANT}
\end{itemize}
The variations specify the colors that are to be used for writing as well as the target grid along with whether or not the writing should be delayed to the end of the turn. In the case of the \texttt{WRITE}, \texttt{WRITE TO MEMORY}, \texttt{WRITE INSTANT} and \texttt{WRITE TO MEMORY INSTANT}, the colors to be written are the ones specified in the command grid.

\subsection{\texttt{CHECK}}
The \texttt{CHECK} command compares the color contents of the target grid with the source ones and if they do not match, terminates the current rule, does nothing otherwise. Light green source tiles are skipped and as such, if all the source tiles are light green, the command is equivalent to an \texttt{if true} instruction. These are the command variations:
\begin{itemize}
    \item \texttt{\texttt{CHECK}}
    \item \texttt{CHECK NOT}
    \item \texttt{CHECK MEMORY}
    \item \texttt{CHECK MEMORY NOT}
    \item \texttt{CHECK WITH MEMORY}
    \item \texttt{CHECK WITH MEMORY NOT}
\end{itemize}
The source tiles are the command tiles if the command type is \texttt{CHECK}, \texttt{CHECK NOT}, \texttt{CHECK MEMORY} or \texttt{CHECK MEMORY NOT}. Otherwise, the source tiles are the memory tiles. The \texttt{NOT} variations compare the source and target tiles and terminate the current rule if they match.
When the source tiles are not the command tiles themselves, the command tile colors indicate the referenced tile index. 

\subsection{Color Enumeration Mapping}
\label{sec:colenum}
Specifically, the colors in Figure~\ref{fig:colors} are transformed into numbers from \texttt{1} to \texttt{9} and mapped onto the source tile enumerations, which are equivalent to the positional command enumeration in Figure~\ref{fig:mekrule}.

\subsection{\texttt{SHIFT}}
The \texttt{SHIFT} command translates the focus location in all, within the command specified, directions. Specifically, the operation of this command is as follows: if a tile is colored dark green (the only alternative color for this command), the subsequent commands are executed with the focus shifted by one tile in the direction of the coloured neighbour relative to the center tile. 

\begin{figure}[h]
\centering
\includegraphics[width=\linewidth]{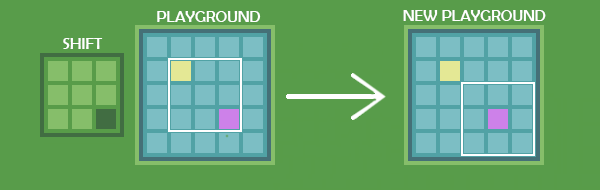}
\caption{\texttt{SHIFT} command: the focus moves from the center of the playground, one step in the south-east direction, because of the dark green tile location in the \texttt{SHIFT} command grid.}
\label{fig:shift}
\end{figure}

Consider Figure~\ref{fig:shift}. On the left, the \texttt{SHIFT} command (dark green outline) is shown with the bottom-right tile in dark green. Next to it is the playground. The white \texttt{3x3} rectangle outline represents the focus location before the command execution. The playground on the right shows the focus location after the command execution.

If multiple tiles are dark green, the execution of the subsequent commands repeats for each direction, resetting the pre-shift focus location before every shift. If the center tile is colored or no tiles are colored, the following commands are executed without shifting. 

\subsection{\texttt{ROTATE}}
Similarly to the \texttt{SHIFT} command, the \texttt{ROTATE} command executes the subsequent commands with the focus rotated. Specifically, dark green command grid tiles indicate the amount by which the subsequent commands should be rotated clock-wise around the center tile. Coloring the top center tile applies no rotation, the top-right one has the commands rotate by one tile clock-wise (45 degrees), the right tile has them rotate by two tiles and so on. If the center tile is dark green colored or no tiles are colored, the subsequent commands are executed without rotation. 

\begin{figure}[h]
\centering
\includegraphics[width=\linewidth]{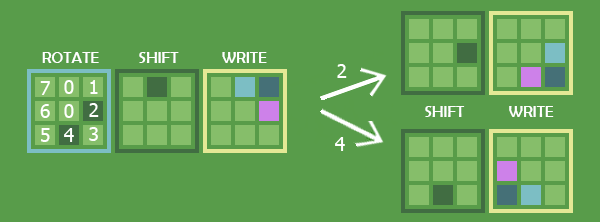}
\caption{\texttt{ROTATE} command: generates multiple rules from the following commands, which are all rotated clock-wise based on the dark green command tile index.}
\label{fig:rotate}
\end{figure}

Consider Figure~\ref{fig:rotate}. Here, a three-command rule on the left starts with a \texttt{ROTATE} command (light blue command border), followed by a \texttt{SHIFT} and \texttt{WRITE}. When the rule is executed, the commands executed are those seen on the right -- for each dark green colored tile, a row of rotated subsequent commands is generated. This forking also happens when multiple tiles in the \texttt{SHIFT} command are colored dark green.

\subsection{\texttt{CALL}}
The \texttt{CALL} command is used to indicate when another rule should be called. Executing a \texttt{CALL} command, executes the rule indicated by the dark green tile index. Upon completion of the rule, it returns to continue with the commands following the \texttt{CALL} command, using the same indication as the rule index grid (Figure~\ref{fig:rule_indexes}). Only one dark green colored tile is allowed to be used in the command. For example, the \texttt{CALL} command in Figure~\ref{fig:rule} would call rule $5$.

\subsection{\texttt{CYCLE}}
The \texttt{CYCLE} command cycles the target tiles the same way a click does on a command tile. The tiles are cycled by the amount equal to the index of the command color (see Section~\ref{sec:colenum}). For example, if the center command tile is colored light blue, the center of the focus tiles will be cycled by \texttt{2} colors (the light blue color index is \texttt{2}). These are the command variations:
\begin{itemize}
    \item \texttt{CYCLE}
    \item \texttt{CYCLE INSTANT}
    \item \texttt{CYCLE MEMORY}
    \item \texttt{CYCLE MEMORY INSTANT}
\end{itemize}


\section{Limitations}
\subsection{MekLang}
Even though the introduced language is flexible enough to implement a range of distinct recognizable mechanics and games, the flexibility is nowhere near that of a general purpose language. That is, the cost of making game mechanics prototyping more accessible is the loss of access to established tools and libraries. Furthermore, the poor performance of the language is also a considerable drawback. Depending on the complexity of the logic implemented, the logic may take a while to compute. A while here is considered  up to a second, making rapid changes to the board state unobtainable. 

\subsection{Interface}
Figure~\ref{fig:mek} shows the interface of Mek, with all the described parts labeled. This initial implementation limits the number of rules and commands available to the user, primarily due to the visual representation restrictions. This inhibits and biases the designer creativity most significantly. A good example for this is the inability to implement the ko-rule of \emph{Go}, which requires remembering all the distinct game states that have already occurred. This example could easily be implemented given multiple playgrounds and the ability to select the targeted one.

\subsection{Game Making}
Moreover, making games in this initial implementation is considerably more difficult compared to making individual mechanics. Unlike the previous limitations, this one appears by design -- the language is focused on allowing rapid mechanic iteration. The only reason games (set of mechanics with an end condition) can be made in it is because end conditions are mechanics too. The impact of the described limitations can also still be reduced, which is the primary goal of the future work.

\section{Conclusion and Further Work}
\label{sec:futer}
In this document, a language for prototyping mechanics for two-dimensional, turn-based, tile-based, complete-information games was introduced. The language was described in detail, shown to be capable of implementing a range of distinct mechanics (and games) and the limitations of the implemented system were outlined. 

With the ability to succinctly describe a range of mechanics, the opportunities to more easily explore them open up. Specifically, the future work is laid out to be the the implementation of additional features, which allow automating the design of the mechanics and comparing the mechanics statistically in parallel with reducing the friction of using the language for its main purpose -- prototyping mechanics.

The additional features are expected to be: the removal of the grid size limitations (both of the playground and the commands), support for API-level access through a webserver, ability to setup multiple playgrounds as well as the support for custom images instead of preset range of colors.

\section{Demo}
A working demonstration with all the mentioned implemented mechanic setups can be found on the author's website\footnote{https://rokasv.com/mek-paper-1/}. The tool shown there is a demonstration of the described concepts and is not yet intended for professional use.

\section*{Acknowledgment}
This work was funded by the EPSRC CDT in Intelligent Games and Game Intelligence (IGGI) EP/L015846/1.

\bibliographystyle{unsrt}
\bibliography{refs}

\end{document}